\documentclass[sn-mathphys-num]{sn-jnl}%

\usepackage{graphicx}%
\usepackage{multirow}%
\usepackage{amsmath,amssymb,amsfonts}%
\usepackage{amsthm}%
\usepackage{mathrsfs}%
\usepackage[title]{appendix}%
\usepackage{xcolor}%
\usepackage{textcomp}%
\usepackage{manyfoot}%
\usepackage{booktabs}%
\usepackage{algorithm}%
\usepackage{algorithmicx}%
\usepackage{algpseudocode}%
\usepackage{listings}%
\usepackage{lineno}
\theoremstyle{thmstyleone}%

\theoremstyle{thmstyletwo}%

\theoremstyle{thmstylethree}%

\begin{document}
	
	\title[Article Title]{The interplay between liquid-liquid and ferroelectric phase transitions in supercooled water}

	\author*[1,2]{\fnm{Maria Grazia} \sur{ Izzo}}\email{mariagrazia.izzo@unive.it; mariagraziaizzo@gmail.com}
	
	\author[3]{\fnm{John} \sur{Russo}}\email{john.russo@uniroma1.it}
	
	\author[4]{\fnm{Giorgio} \sur{Pastore}}\email{pastgio@units.it}

	\affil[1]{\orgdiv{Dipartimento di Scienze Molecolari e Nanosistemi}, \orgname{Università Ca' Foscari Venezia}, \orgaddress{\street{Via Torino 155}, \city{Venezia Mestre}, \postcode{30172},  \country{Italy}}}
	
	\affil[2]{\orgname{Scuola Internazionale Superiore di Studi Avanzati SISSA}, \orgaddress{\street{via Bonomea, 265}, \city{Trieste}, \postcode{34136}, \country{Italy}}}
	
	\affil[3]{\orgdiv{Dipartimento di Fisica}, \orgname{Sapienza Università di Roma}, \orgaddress{\street{Piazzale Aldo Moro, 5}, \city{Roma}, \postcode{00185}, \country{Italy}}}
	
	\affil[4]{\orgdiv{Dipartimento di Fisica}, \orgname{Università degli Studi di Trieste}, \orgaddress{\street{Via Valerio 2}, \city{Trieste}, \postcode{34127}, \country{Italy}}}

	\abstract{The distinctive characteristics of water, evident in its thermodynamic anomalies, have implications across disciplines from biology to geophysics. Considered a valid hypothesis to rationalize its unique properties, a liquid-liquid phase transition in water below the freezing point, in the so-called supercooled regime, has nowadays been observed in several molecular dynamics simulations and is being actively researched experimentally. The hypothesis of ferroelectric phase transition in supercooled water can be traced back to 1977, due to Stillinger. In this work, we highlight intriguing and far-reaching implications of water: the ferroelectric and liquid-liquid phase transitions can be designed as two facets of the same underlying phenomenon. Our results are based on the analysis of extensive molecular dynamics simulations and are explained in the context of a classical density functional theory in mean-field approximation valid for a polar liquid, where dipolar interaction is treated perturbatively. The theory underpins the potential role of ferroelectricity in promoting the liquid-liquid phase transition, being the density-polarization coupling inherent in the dipolar interaction potential. The existence of ferroelectric order in supercooled low-density liquid water is confirmed by the observation in molecular dynamics simulations of collective modes in space-time polarization correlation functions, traceable to spontaneous breaking of continuous rotational symmetry. Our work sheds light on water’s supercooled behavior and opens the door to new experimental investigations of the static and dynamic behavior of water’s polarization.}
	
	
	\maketitle
	
	The identification of equilibrium water's thermodynamic anomalies, notably density ($\rho$) maximum at 277 K, compressibility, and specific heat minima around  310 K and 280 K, respectively \cite{Angell}, has immediately sparked broad scientific interest.
	A first-order liquid-liquid phase transition (LLPT) between a high-density liquid (HDL) and a low-density liquid (LDL) in the supercooled regime, was proposed \cite{Poole} to explain equilibrium water's anomalies and polyamorphism \cite{Bellisent-Funel}. The first observation of HDL and LDL water in molecular dynamics (MD) simulations is reported in Ref. \cite{Poole}. Recently, extensive MD simulations in realistic \cite{Debenedetti}, as well as \textit{ab initio} neural network \cite{Gartner}, models of water have clearly supported the first-order LLPT existence.
	The first-order LLPT line ends at a second-order critical point (CP). Despite experimental hints \cite{Kim2}, direct evidence is challenging due to water's crystallization tendency near the MD simulations-predicted CP.
	The Widom line (WL) \cite{Xu}, located in the pressure-temperature thermodynamic plane ($p$, $T$) region above CP but yet in the supercooled state, has been observed via both MD simulations \cite{Xu} and experiments \cite{Kim1}. Crossing the WL from high $T$, water transforms smoothly from HDL-like to LDL-like configurations \cite{Xu,Kim1}, while the isothermal compressibility ($K_T$) reaches a local maximum \cite{Sciortino1}.
	Beyond $\rho$, different order parameters have been proposed to characterize the LLPT and elucidate its physical origin, mostly based on local structure geometry \cite{Oca,Tanaka,Foffi2}. Recent insights, furthermore, indicate that varying degrees of topological order of hydrogen-bond network can distinguish HDL from LDL \cite{Neophytou}.
	
	The hypothesis of ferroelectric phase transition in supercooled water stems from a 1977 proposal by Stillinger \cite{Stillinger}, following the observation of proton ordering in certain ice polymorphs \cite{Eisenberg}.
	During the same years, measurements of the dielectric constant of supercooled water emulsions at ambient $p$ down to T=238 K \cite{Hasted, Hodge} revealed an increase in dielectric constant as $T$ decreases. Ref. \cite{Hodge} emphasized that this trend aligns with divergence at 228 K, close to the WL \cite{Sciortino1,Kim1}, albeit with a rather small critical exponent.
	Although the idea persisted over the years - Refs. \cite{Shelton, Pounds} explore the potential ferroelectricity of equilibrium water - the connection between ferroelectricity and LLPT in supercooled water was not examined.
	A series of papers \cite{Menshikov, Fedichev1, Fedichev2} deal with modeling the free energy of supercooled water in light of a possible ferroelectric phase transition near the WL. The ferroelectric and the liquid-liquid phase transitions were, however, always considered as two distinct and concomitant phenomena. By comparing the expression of the free energy presented in Ref. \cite{Menshikov} with the one provided in Eq. \ref{Gibbs_ansatz}, it becomes clear that the former lacks the density-polarization coupling term, thus undermining the foundational premise to relate ferroelectric and LLPT. In Ref. \cite{Fedichev2}, foer example, supercooled water was assumed to be a mixture of HDL and LDL, with only LDL presumed to undergo a ferroelectric phase transition. As further support of this, the hypothesis that a ferroelectric phase transition could also occur at the first-order LLPT line has never been considered.
	Beyond ices polymorphs \cite{Eisenberg}, incipient ferroelectricity has been observed in confined water \cite{Gorshunov}. Interestingly, electrofreezing of water at ambient conditions was observed recently by \textit{ab initio} MD simulations, highlighting the existence of an amorphous ferroelectric ice \cite{Cassone}. On the other hand, dielectric measurements reveal distinct differences in the dielectric properties of high- and low-density amorphous ices (HDA/LDA) \cite{Andersson}. Extrapolated from Ref. \cite{Andersson}, LDA and HDA dielectric constant is respectively $\sim 10$ at $T \sim$ 130 K and $p$=4000 bars, and $\sim$ 130 at same $T$ and $p$=6000 bars.
	
\begin{figure*}[t!]
		\centering
		\includegraphics[width=1\linewidth]{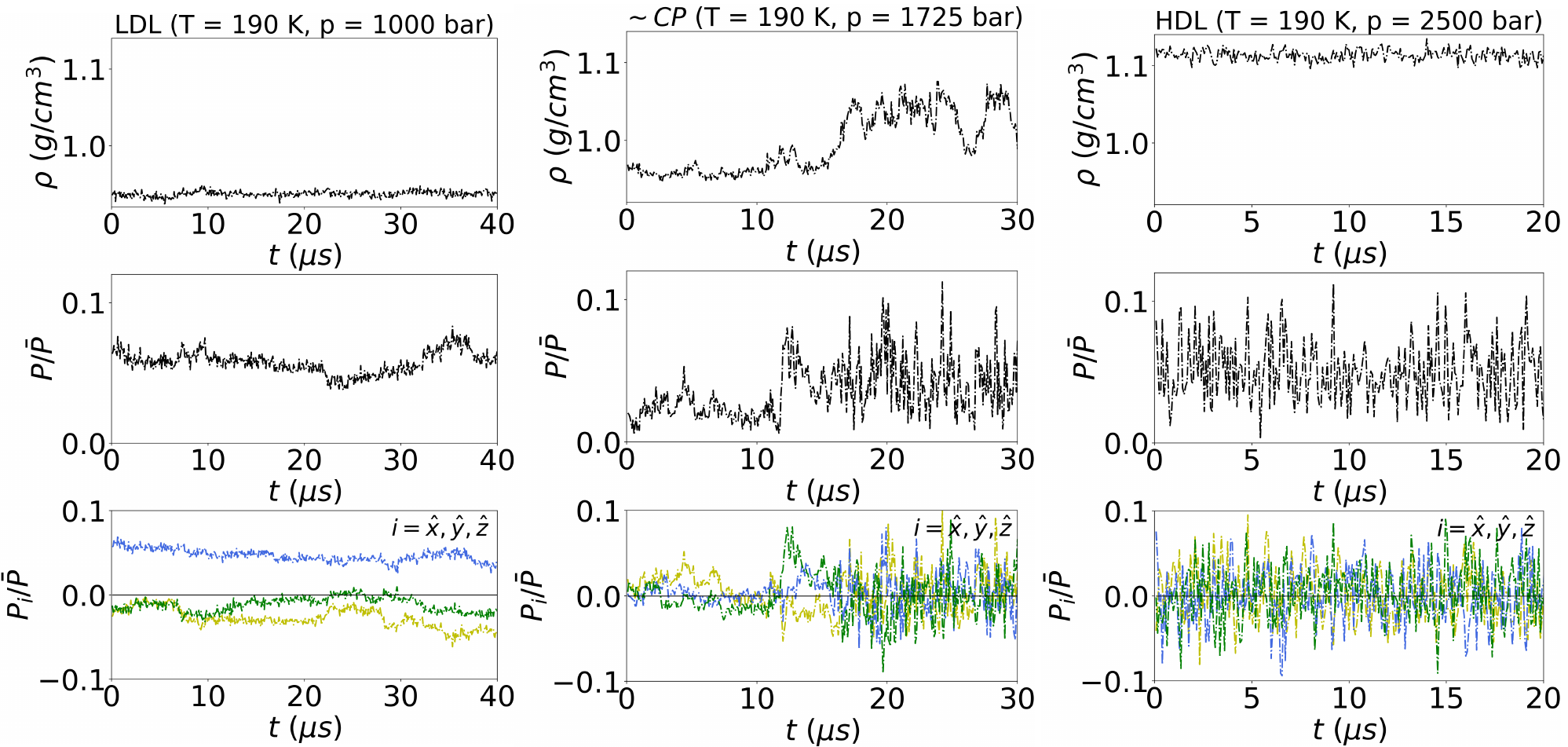}
		\caption{Temporal evolution of $\rho$ (top) $P$ (middle) and $P_i$ (bottom) across supercooled water in LDL (left), near CP (center) and in HDL (right) as obtained from MD simulations, suggesting paraelectric and ferroelectric character for HDL and LDL, respectively. It is $\bar{P}=N d$. The CP for TIP4P/Ice model has been evaluated \cite{Debenedetti} to be $( \bar{p}_c=1725 \ bar, \bar{T}_c=188.6 \ K)$. }\label{Fig2short}
	\end{figure*}

	Reanalyzing the water MD simulations of Ref. \cite{Debenedetti}, a distinct correlation between $\rho$ and total polarization magnitude ($P$) emerges: while HDL retains paraelectric characteristics, the trend of LDL polarization suggests a ferroelectric character, as shown in Fig. \ref{Fig2short}. A qualitatively similar result was obtained recently in Ref. \cite{Malosso}, where simulations based on \textit{ab initio} deep neural-network force field are presented. Thereby this result is resilient to changes in the MD simulation's potential and water model. The persistence of the result when moving from an \textit{ab initio} deep neural-network force field, which includes molecular polarizability, to empirical potentials with rigid non-polarizable molecules, indicates that the primary effect is due to the orientation of molecular dipoles rather than molecular polarizability. This provides a more solid foundation for our subsequent developments. Though it clearly shows the existence of a correlation between $P$ and $\rho$, this result does not demonstrate what is the role of $P$ in the LLPT.
	Since the dipolar degrees of freedom drawing P, though coupled to positional degrees of freedom, are governed by a different interaction potential, occurrence that distinguishes $P$ from other proposed order parameters \cite{Oca,Tanaka,Foffi2}, the hypothesis can be advanced that $P$ plays an active role, different from that of $\rho$, in the LLPT.
	An analogy akin to that of ferroeleastic phase transition in crystals \cite{Landau, Chaikin} or liquid crystals \cite{Sebastian} can be envisioned.  
	\begin{figure*}[t!]
		\centering
		\includegraphics[width=1\linewidth]{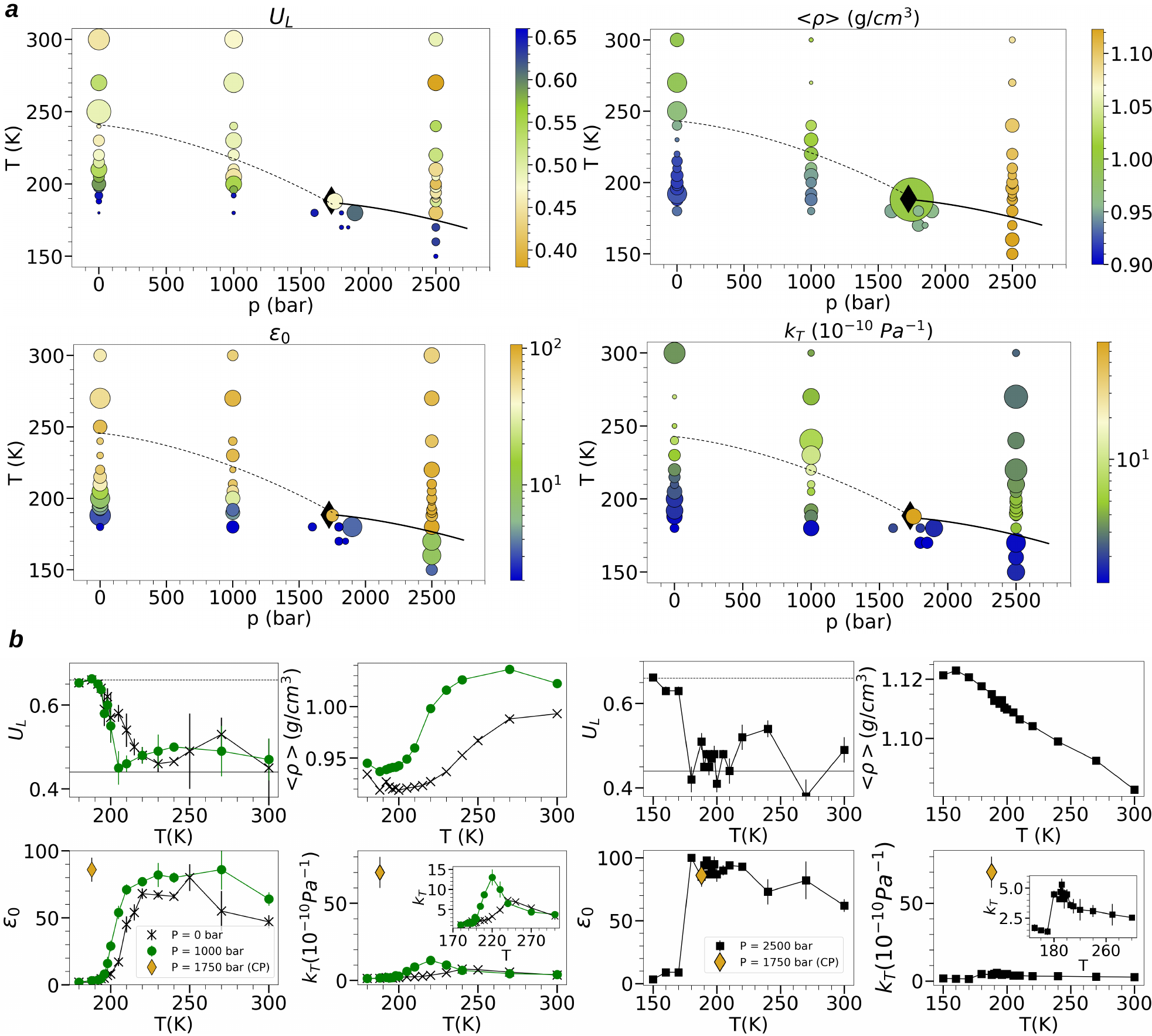}
		\caption{\textbf{\textit{a}}, The figure depicts $U_L$, $\rho$, $\epsilon_0$, and $K_T$ at different points in the (T, p) plane obtained from MD simulations. Color scale represents the quantities value, symbol size is proportional to the associated error, obtained through block averaging. Full and dashed black lines serve as visual guidance, marking the first-order LLPT and WL, respectively. The diamond symbol marks CP. \textbf{\textit{b}}, $U_L$, $\langle\rho\rangle$, $\epsilon_0$, and $K_T$ as functions of $T$ along constant-$p$ lines intersecting the WL ($p$=0  bar, 1000 bar, left) and the first-order LLPT line ($p$=2500 bar, right). The values of each quantity at CP are marked by a diamond symbol. A gradual change is observed in $U_L$ and $\rho$ when crossing the WL, while a sudden shift in $U_L$ occurs at the first-order LLPT line.}\label{Fig3short}	
	\end{figure*}
	On this trail, we firstly obtain from MD simulations the $P$-$\rho$ phase diagram in the ($p$, $T$) plane, secondly, starting from the microscopic interaction potential and treating the dipole interaction perturbatively, we develop a classical density functional theory (DFT) in a mean-field approximation. Unlike previous treatments, such as the notable examples in the Stockmayer fluid \cite{Groh}, our theory uniquely features the emergence of coupling between $P$, the order parameter, and $\rho$, in the free energy, stemming directly from the dependence of the dipolar interaction potential from the positional degrees of freedom.  Interestingly, Ref. \cite{Kornyshev} proposes that the coupling between P and $\rho$ fluctuations can explain some of the non-local dielectric properties of water. However, it should be kept in mind that we are dealing with their macroscopic counterparts here, and this coupling cannot explain the emergence of a connection between ferroelectricity and LLPT. Our developments aim rationalizing the LLPT in water, modeled as a dipolar liquid. It does not exclude that different systems or different water models with different microscopic interactions \cite{Chan} could led to the same phenomenology. A similar DFT approach could, in these cases, be applied.
	\section{Results} 
	\subsection{Ferroelectric character of the LLPT in supercooled water}
	\label{subsec2_1}
	Fig. \ref{Fig2short} shows the temporal evolution of $P$, polarization components along the three spatial directions ($P_i$) and $\rho$ obtained by reanalyzing extensive MD simulations of TIP4P/Ice water lasting up to 40 $\mu$s of Ref. \cite{Debenedetti}. It emerges a clear correlation between $P$, $P_i$ and $\rho$ trends. MD simulations employed \textit{isothermal-isobaric} (NpT) ensemble with $N=1000$ molecules. For additional details on MD simulations and analysis, refer to \textit{Materials and Methods}, Sec. \ref{Met1}, \ref{Met2} and Ref. \cite{Debenedetti}. In the LDL phase, all $P_i$'s maintain a non-zero value, while in the HDL phase, they show large fluctuations around a zero mean value. At CP = $(\bar{p}_c,\bar{T}_c)$, where the temporal series shows the characteristic bimodal behavior, the $\rho$ fluctuation between LDL and HDL coincides with a transition in $P$ and $P_i$'s trend. Analyses covering several ($p$, $T$) points are in \textit{SI} Appendix, Sec. I. 
	While there is never spontaneous magnetization in a finite system, the ferroelectric phase exhibits a nonzero $<P_i>$, as in Fig. \ref{Fig2short}, as long as the box size $L$ of MD simulations exceeds the correlation length of the order parameter. The brackets $< \ >$ denote ensemble average. Nevertheless, obtaining a full characterization of the polarization probability distribution is challenging because of spatial domains with varying polarization and orientation transitions. 
	The $4th$ order cumulant of the polarization probability distribution, so-called Binder cumulant $U_L=1-\frac{1}{3}\frac{<P^4>}{<P^2>^2}$, has emerged as a powerful tool for discerning between paraelectric and ferroelectric phases \cite{Binder, Holm, Weis}. 
	We derived $U_L$, $<\rho>$, the dielectric constant, $\epsilon_0$, and $K_T$ from MD simulations, as detailed in \textit{Materials and Methods}, Sec. \ref{Met2}. 
	It is $\epsilon_0=1+\chi$, where $\chi$ is the electric susceptibility. These quantities are shown in Fig. \ref{Fig3short}, supporting that the LLPT and the behavior at WL in supercooled water involves the dipolar degrees of freedom, as discussed below. (i) $U_L$ passes from about $4/9$, indicative of a paraelectric phase \cite{Holm}, to about $2/3$, indicative of a ferroelectric phase \cite{Holm, Binder}, crossing from high-$T$ both the WL (smooth transition) and the first order LLPT line (sharp transition).
	(ii) $<\rho>$ gradually decreases as it crosses the WL from high-$T$, transitioning from typical values of HDL to LDL. The transitions of $<\rho>$ and $U_L$ co-occur. The $\rho$-transition expected when crossing the first-order LLPT line is obscured by the $<\rho>$-variation with $T$ at those p's.
	(iii) Crossing the WL or the LLPT line from low-$T$, $\epsilon_0$ increases to reach its maximum along the WL and the first-order LLPT line. (iv) $K_T$ exhibits a maximum along the WL and the first-order LLPT line. (v) Along the WL, the maximum value of $K_T$ increases as it approaches $\bar{p}_c$, where it reaches a large value consistent with a divergence. Notably, unlike $K_T$, the peak value of $\epsilon_0$, still reaching large values, remains almost unchanged along the WL and at CP. Considering that $K_T$ and $\epsilon_0$ are related to the second derivative of the free energy with respect to $\rho$ and $P$, respectively, this result suggests that the features of the free energy as a function of $P$ or $\rho$ differ. Fig. \ref{Fig2short} shows the existence of a correlation between $\rho$ and $P$. However, the results in points (i)-(v) provide additional insights. If $P$ were simply coupled to $\rho$ by a linear relationship, $P$ and $\rho$ could be used interchangeably as the order parameter of the LLPT. However, this is not the case here because if it were, $K_T$ and $\chi$ would exhibit the same trend.
	
	In \textit{SI} Appendix, Sec. II, the local spatial distribution of dipoles in LDL and HDL is presented. A complete characterization is beyond the scope of this manuscript, which focuses on the emergence of spontaneous macroscopic polarization in LDL. However, the potential appearance in LDL of a local dipolar order resembling a chiral pattern, in line with the molecular chiral order observed in Ref. \cite{Matsumoto}, warrants further investigation.
	To gain further and complementary insights into microscopic spatial distribution of masses and dipoles, \textit{SI} Appendix, Sec. III shows the wavevector ($\textbf{k}$)-dependent transverse and longitudinal to $\textbf{k}$ static dielectric functions, respectively $\epsilon_{T\hat{k}}(k)$ and $\epsilon_{L\hat{k}}(k)$, and the  static structure factor, $S(k)$, in the HDL, LDL and close to CP. $\textbf{k}=k \hat{k}$ is the Fourier conjugate variable of the space variable $\textbf{r}$. In this text, bold quantities are vectors, the corresponding non-bold symbols are their magnitudes, and those with a circumflex accent are unit vectors. Averages have been taken over $\hat{k}$, as for all the quantities introduced in the following. Details are in \textit{Materials and Methods}, Sec. \ref{Met2}. 
	The figure in \textit{SI} Appendix, Sec. III reveals divergences followed by negative $\epsilon_L$ values, akin to overscreening phenomena \cite{Fasolino, Bopp}. Interestingly, in Ref. \cite{Kornyshev}  water's overscreening was attributed to $\rho$-$P$ fluctuations coupling.
	\subsection{A classical DFT for the ferroelectric LLPT}\label{subsec2_2}
	In one-component polar liquid, composed of non-polarizable molecules, the number density of particles at the point $\textbf{r}$ having the unit vector $\hat{d}$ as dipole orientation is
	\begin{equation}
		\tilde{\rho}(\textbf{r},\hat{d})=\sum_{i=1}^N \delta(\hat{d}-\hat{d}_i)\delta(\textbf{r}-\textbf{r}_i)=\rho(\textbf{r})\zeta(\Omega,\textbf{r}), \label{density}
	\end{equation}
	where $\textbf{r}_i$ and $\hat{d}_i$ are respectively the center of mass position vector and dipole orientation of the $i$-th particle.
	$\rho(\textbf{r})$ and $\zeta(\textbf{r},\Omega)$ are respectively the particle number density without specified dipole orientation and the probability distribution of dipole orientation at point $\textbf{r}$. $\Omega$ is the solid angle element. As detailed in \textit{SI} Appendix, Sec. IV, introducing molecular polarizability does not qualitatively affect the DFT obtained for non-polarizable molecules.
	To parameterize the Helmholtz free energy in terms of $\tilde{\rho}(\textbf{r})$, we decompose it into two parts: $F_0$, the free energy of a reference system devoid of dipole interaction, and the perturbative term $\mathcal{F}$, which incorporates dipole interaction.
	The structure of most liquids, especially at high density, is indeed primarily influenced by short-range hard-core pair interaction \cite{Hansen}. The perturbative effect will be treated via mean-field approximation, excluding from $\mathcal{F}$ contributions of $\tilde{\rho}(\textbf{r})$ correlations \cite{Hansen}. Assumption of spatial homogeneity fixes $\rho(\textbf{r})=\rho$. We furthermore use for $\zeta(\Omega)$ the simple ansatz
	\begin{equation}
		\zeta(\Omega,\textbf{r})=\zeta(\Omega)=\frac{1+ \boldsymbol{\delta} \hat{d}}{4\pi}, \label{ansatz_zeta}
	\end{equation}
	which can identify paraelectric ($\mathbf{\delta}=0$) and ferroelectric ($\mathbf{\delta}\neq 0$, with total polarization vector $\textbf{P} \propto \mathbf{\delta}$) states, as detailed in \textit{Materials and Methods}, Sec. \ref{Met3}. 
	To facilitate comparison with MD simulations, the NpT ensemble \cite{Hansen} is used. 
	The Gibbs free energy derived from the DFT, with detailed derivation in \textit{Materials and Methods}, Sec. \ref{Met3}, is
	\begin{multline}
		G=\gamma_0(\bar{V})+ \frac{a}{2}(T-T_c)P^2+\frac{B}{4}P^4+\frac{B'}{6}P^6+\frac{M}{2}\Delta V^2- p \beta \Delta V P^2+ p \Delta V-\textbf{E} \cdot \textbf{P}, \label{Gibbs_ansatz}
	\end{multline}
	where $a$, $B$, $B'$, $M$, and $\beta$ are positive constants, $\bar{V}$ is the equilibrium volume of the reference system, $\gamma_0$ is its Gibbs free energy, and $\Delta V$ is the difference between the system's volume ($V$) and $\bar{V}$. $\textbf{E}$ is the external electric field. The potential in Eq. \ref{Gibbs_ansatz} belongs to the class of potentials leading to tricritical points \cite{Chaikin,Patashinskii}, mirroring those of ferroelastic or magnetoelastic crystals \cite{Landau, Chaikin}, where the deformation tensor replaces $p$. Notably, the free energy expression in Eq. \ref{Gibbs_ansatz} and the signs of its coefficients are derived from DFT rather than being arbitrarily chosen to align with the MD simulation results. The key points in the DFT developments, detailed in \textit{Materials and Methods}, Sec. \ref{Met3} are: i) the positional disorder of the liquid which, combined with the microscopic expression of the dipolar potential interaction, leads to the potential cancellation of the coefficient of the $P^2$ term in Eq. \ref{Gibbs_ansatz}; and ii) the $\rho$-Taylor expansion of $\mathcal{F}$ around the density of the reference system $\bar{\rho}$, which, given the characteristics of the dipolar interaction potential, yields the $\rho-P$ coupling term. The  dipolar interaction potential in Eq. \ref{dipoleinter} depends indeed on the vector distance between two dipoles in the liquid. This introduces a dependence of $\mathcal{F}$ on $\rho$ and, through the $\rho$-Taylor expansion, leads to the $\rho-P$ (or $V-P$ in the NpT ensemble) coupling term.
	Note that the sign of the coefficient in front of the coupling term is also determined, beyond the features of the dipolar potential interaction, by positional disorder.  
	The equilibrium values of $\textbf{P}$ and $\Delta V$, $\textbf{P}{eq}$ and $\Delta V{eq}$, are determined through a variational principle,
	\begin{eqnarray}
		&&\left.\frac{\partial G}{\partial \textbf{P}}\right|_{\textbf{P}= \textbf{P}_{eq}}=a(T-T_c)\textbf{P}_{eq}+B\textbf{P}_{eq}^3+B'\textbf{P}_{eq}^5 -2p \beta  \textbf{P}_{eq}  \Delta V_{eq}- \textbf{E}=0; \label{eq_PdensP6_1} \\
		&&\left.\frac{\partial G}{\partial  \Delta V}\right|_{\Delta V=\Delta V_{eq}}=M \Delta V_{eq}-p \beta  P_{eq}^2+p =0.   \label{eq_PdensP6_2}
	\end{eqnarray}
	It follows 
	\begin{equation}
		\Delta V_{eq}=-\frac{p}{M}(1-\beta P_{eq}^2). \label{rho}
	\end{equation} 
	The first term in Eq. \ref{rho} represents the liquid's response to applied $p$, leading to compression, indicated by a negative contribution to $\Delta V_{eq}$. The second term shows that at a given $p$, the system's equilibrium volume $V_{eq}=\bar{V}+\Delta V_{eq}$ is larger when $P_{eq} \neq 0$ than when $P_{eq}=0$. Consequently, a ferroelectric phase exhibits lower density than a paraelectric phase. Eq. \ref{rho} formalizes the presence of a ferroelectric LDL phase and a paraelectric HDL. The sign of the coefficient of the $V-P$ coupling term in Eq. \ref{Gibbs_ansatz}, which is ultimately determined by positional disorder as discussed in \textit{Materials and Methods}, Sec. \ref{Met3}, is essential. It establishes that the HDL is paraelectric and the LDL is ferroelectric, rather than the reverse. It is also noteworthy that the relationship between $\Delta V$ and $P$ established by Eq. \ref{rho} is not linear but quadratic. Beyond the detailed calculations presented in Materials and Methods Sec. \ref{Met4}, this fact ultimately explains the different behaviors of $K_T$ and $\epsilon_0$ along the WL. It demonstrates that, according to DFT and consistently with MD simulations discussed in Sec. \ref{subsec2_1}, $P$ and $\rho$ (or $V$ in the NpT ensemble) are not interchangeable order parameters. 
	We are interested in the possible appearance of spontaneous polarization for $\textbf{E}=0$.
	The basic mechanisms underlying the phase transitions associated with the free energy in Eq. \ref{Gibbs_ansatz} are briefly discussed below. The vanishing of the coefficient in front of the $P^2$ term induces a ferroelectric transition. The emergence of spontaneous polarization then leads, via Eq. \ref{rho}, to a LLPT. By inserting Eq. \ref{rho}, which directly results from the $V-P$ coupling, into Eq. \ref{Gibbs_ansatz}, it becomes evident that the coefficient in front of the $P^4$ term can also vanish, as explicitly shown in \textit{Materials and Methods}. Sec. \ref{Met4}. The simultaneous cancellation of the coefficients of the $P^2$ and $P^4$ terms gives rise to the existence of the tricritical point, whose occurrence remains thus linked to the existence of the $V-P$ coupling.
	The $P$-$\rho$ phase diagram in the ($p$, $T$) plane associated to Eq. \ref{Gibbs_ansatz} is obtained by analyzing the stability of solutions, Eqs. \ref{eq_PdensP6_1} and \ref{eq_PdensP6_2}, set by the conditions $\chi, K_T >0$. See \textit{Materials and Methods}, Sec. \ref{Met4} and \textit{SI} Appendix, Sec. V for detailed discussions. A schematic representation is shown in Fig. \ref{Fig1short}, with comments below.
	
	\begin{enumerate}
		\item
		\textit{Ferroelectric (first-order) LLPT.} For constant $p>\bar{p}_c$ at $T=\hat{T}_c(p)$ a first-order ferroelectric LLPT is predicted. The following points support evidence for the first-order nature. (i) By lowering $T$ until $\hat{T}c$, $P$ switches abruptly from $P{eq}=0$ to $P_{eq} \neq 0$ (see Tab. \ref{tab:table4} in Materials and Methods Sec. \ref{Met4}), and  consequently $V_{eq}=\bar{V}-\frac{p}{M}$ shifts to $V_{eq}=\bar{V}-\frac{p}{M}(1-\beta P^2_{eq})$.
		(ii) There exists a $T$-range above and below $\hat{T}_c$ where the ferroelectric LDL and the paraelectric HDL, respectively, are metastable (see Tab. \ref{tab:table4} in \textit{Materials and Methods}, Sec. \ref{Met4}). (iii) Along the curve $T=\hat{T}_c(p)$ the two phases are both stable and coexist. (iv) At $T=\hat{T}_c(p)$, neither the $\chi$ nor $K_T$ diverges, but both reach a local maximum. 
		(v) At $(\bar{T}_c,\bar{p}_c)$, the end point of the first-order phase transition, the $P$- and $V$-difference between the two phases goes to zero, while $\chi$ and $K_T$ diverge. 
		Comparing with MD simulations in Sec. \ref{subsec2_1} identifies $T=\hat{T}_c(p)$ as the first-order LLPT line.
		\begin{figure}
			\centering
			\includegraphics[width=0.6\linewidth]{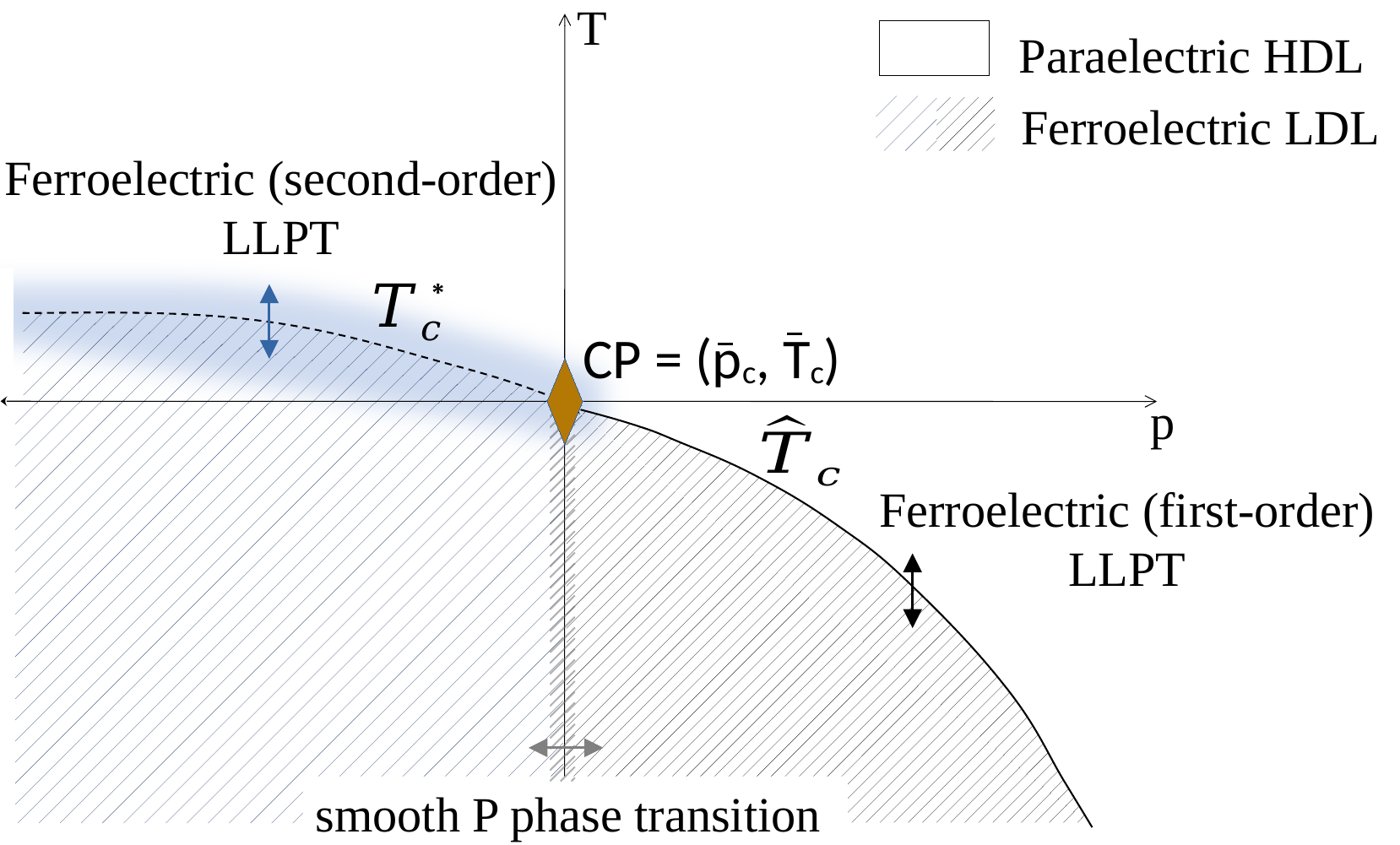}
			\caption{Pictorial representation of the $P-\rho$ phase diagram of the polar liquid in the $($p$, $T$)$ plane obtained from mean field DFT. $CP$ is marked by a diamond symbol. For $p<\bar{p}_c$ a first-order LLPT is predicted at $T=\hat{T}_c(p)$, represented by a full black line. For $p>\bar{p}_c$ a second-order ferroelectric LLPT is predicted at $T=T^*_c(p)$, marked by a black dashed line bordered by blue nuance. For $T<\bar{T}_c$ the theory predict a ferroelectric phase. A smooth transition in the $P$ value is predicted upon crossing the line $p=\bar{p}_c$. The two ferroelectric phases with different $P$ values are distinguished in the figure by varying degrees of texture's line density.}\label{Fig1short}
		\end{figure}
		\item 
		\textit{Ferroelectric (second-order) continuous LLPT.} 
		At constant $p<\bar{p}_c$, the theory predicts a second-order ferroelectric phase transition at $T=T_c^*(p)$. The order parameter $P$ increases continuously from zero for $T<T_c^*$ (see Tab. \ref{tab:table2} in \textit{Materials and Methods}, Sec. \ref{Met4}). $V$ decreases continuously according to Eq. \ref{rho}, indicating a simultaneous smooth $\rho$-transition.
		At $T=T_c^*(p)$, $\chi$ diverges. Relevantly, however, the theory predicts $K_T$ reaches a maximum rather than diverging at $T_c^*$. Along the curve $T=T_c^*(p)$, $K_T$ increases with increasing $p$ until it diverges at $\bar{p}_c$.
		Details are in \textit{Materials and Methods}, Sec. \ref{Met4}. 
		Though a finite scaling analysis is required to confirm $\epsilon_0$ divergence in MD simulations, the theory accurately predicts the $K_T$, $U_L$ and $<\rho>$ trend at the WL. This consistency supports identifying the WL with the curve $T=T_c^*(p)$.
		\item 
		\textit{Smooth polarization magnitude phase transition.} For $T<\bar{T}_c$, the system exhibits a ferroelectric phase. However, crossing the line $p=\bar{p}_c$ at a given $T<\bar{T}_c$ with increasing $p$, $P$  gradually increases and V decreases (refer to Tabs. \ref{tab:table4} and \ref{tab:table2} in \textit{Materials and Methods}, Sec. \ref{Met4}). No singular behavior occurs in $\chi$ or $K_T$.
		Observing this transition in MD simulations is challenging due to the overlapping changes in $\Delta V$ and $\bar{V}$ when $p$ increases at fixed $T$. Also, measuring $P$ in MD simulations is demanding, as a non-zero average value is strictly only apparent in the macroscopic limit. Thus, we haven't pursued this.
	\end{enumerate}
	In the proposed theoretical framework, the CP in supercooled water is identified as a tricritical point, leading to critical behavior distinct from that of an ordinary second-order phase transition in the three-dimensional Ising model. According to mean field theory, a second-order phase transition occurs at the tricritical point, but with distinctive critical exponents $\beta$ and $\delta$, which describe the behavior of the order parameter as T approaches the critical T from below and for infinitesimal changes in the conjugate field. For a tricritical point, $\beta = 1/4$ and $\delta = 5$, while for an ordinary second-order phase transition, $\beta = 1/2$ and $\delta = 3$ \cite{Chaikin}. Experimental or numerical determination of these critical exponents at CP could validate the tricritical phase diagram scenario. Note that the order parameter referred to here is $P$, for which the coefficients of the second- and fourth-order terms in the $G$ expansion vanish at the triciritcal point. Since from Eq. \ref{rho} $\Delta \rho$  is proportional to $P^2$, the behavior of $\Delta \rho$ near CP as $T\rightarrow \bar{T}_c^-$, in the tricritical phase diagram scenario would result in $\beta=1/2$, matching that of the three-dimensional Ising model. Interestingly, monitoring the dependence of $P$ on a weak electric field $E$ at CP to determine $\delta$ could help validate the tricritical phase diagram scenario.
	
	Our elementary theory is not conceived to realistically determine the critical $T$'s and $p$'s as these may depend on microscopic details, as well as molecular polarizability, that it overlooks. Instead, it demonstrates that the $P$-based tricritical point scenario is qualitatively supported by a polar liquid, aiming to identify the key mechanisms underlying the LLPT in water. A similar rationale supports using the classical TIP4P/ice potential in MD simulations.
	\subsection{Ferroelectric order in LDL}
	A ferroelectric phase is characterized by the spontaneous breaking of the continuous rotational symmetry group $O(3)$, leading to distinctive behaviors in both the static and dynamic correlation functions of $\textbf{P}$-fluctuations in the direction transverse and parallel to $\hat{p}=\textbf{P}_{eq}/P_{eq}$, $\delta P_{T\hat{p}}$ and $\delta P_{L\hat{p}}$, respectively. A detailed discussion is presented in \textit{SI} Appendix, Sec. VI. If the identification of LDL as a ferroelectric phase is correct, the correlation functions obtained by MD simulations in LDL will thus uncover such features. The following analysis aims to verify this. In \textit{Materials and Methods}, Sec. \ref{Met2} it is described how the correlation functions are obtained from MD simulations. One of the main signs of spontaneous rotational symmetry breaking is the divergence of the $k$-dependent static susceptibility of the so-called symmetry-restoring variable \cite{Forster}, that is $\delta P_{T\hat{p}}$ in the case of ferroelectricity, as $k^{-2}$ in the macroscopic limit $k \rightarrow 0$. The static susceptibility of the transverse to $\hat{p}$ polarization fluctuations, $\chi_{T\hat{p}}(k)$, which represents the space correlation function of $\delta P_{T\hat{p}}$ at time $t=0$, is thus expected to have this trend. Furthermore, so-called  Goldstone modes \cite{Forster} emerge in $C_{T\hat{p}T\hat{p}}(k,t)$, the wavevector and time dependent correlation function of $\delta P_{T\hat{p}}$, representing its space and time correlations. Goldstone modes are propagating, leading to oscillations in time in $C_{T\hat{p}T\hat{p}}(k,t)$, with a characteristic frequency that depends on $k$, meaning the dispersion relation is not constant.
	Detailed analysis is presented in \textit{SI} Appendix, Sec. VI in the framework of the Mori-Zwanzig memory function formalism \cite{Forster}.

	Fluctuations in $\hat{p}$-longitudinal polarization, $\delta P_{L\hat{p}}$, can arise from $P$-fluctuations, $\delta P$, whose dynamics is empirically described by the Landau-Khalatnikov-Tani equation \cite{Tang, Widom}.
	This predicts collective modes, also known as Higgs-like modes \cite{Prosandeev}, in the correlation function $C_{L\hat{p}L\hat{p}}(k,t)$ of $\delta P_{L\hat{p}}$, exhibiting propagating behavior with a linear dispersion relation at moderately small $k$ values \cite{Tang}, changing to constant as $k \rightarrow 0$.
	Goldstone-like and Higgs-like modes can coexists \cite{Prosandeev}.
	 
	Unlike what might be expected, a coupling between $\delta P_{T\hat{p}}$ and $\delta P_{L\hat{p}}$ is possible. It can arise from the constant-modulus principle \cite{Patashinskii}, which establish that lowest order fluctuations satisfy the condition $\delta P^2 = 2P_{eq}\delta P_{L\hat{p}} + \delta P_{T\hat{p}}^2 = 0$ \cite{Patashinskii}. The onset of a polarization fluctuation in the direction transverse to $\hat{p}$ will induce a fluctuation in the direction longitudinal to $\hat{p}$ to preserve $P$ constant, and vice versa.
	\begin{figure*}[t!]
		\centering
		\includegraphics[width=0.7\linewidth]{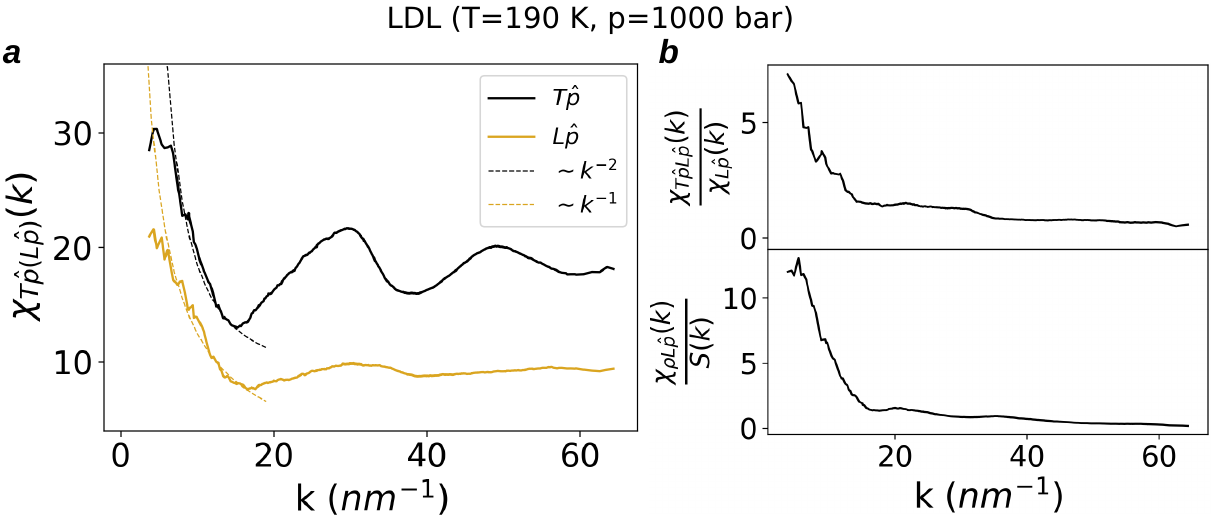}
		\caption{\textbf{\textit{a}}. Static susceptibilities $\chi_{T\hat{p}(L\hat{p})}(k)$, in LDL as a function of $k$. Both $\chi_{T\hat{p}}$ and $\chi_{L\hat{p}}$ show significant enhancement as $k \rightarrow 0$, following trends approximately proportional to $k^{-2}$ and $k^{-1}$, respectively, as indicated by the dashed lines. This behavior is consistent with predictions for a ferroelectric phase characterized by the spontaneous breaking of the $O(3)$ symmetry group.
			\textbf{\textit{b}}. The upper graph shows the static correlation between $\delta P_{L\hat{p}}$ and $\delta P_{T\hat{p}}^2$, represented by $\chi_{T\hat{p}L\hat{p}}$, over $\chi_{L\hat{p}}(k)$ in the LDL phase. The lower graph depicts the static correlation between $\delta \rho$ and $\delta P_{L\hat{p}}$, $\chi_{\rho L\hat{p}}(\textbf{k})$ over $S(k)$ in LDL. A correlation between $\delta P_{L\hat{p}}$, $\delta P_{T\hat{p}}^2$ and $\delta \rho$ emerges, in particular at moderately small $k$s.}\label{Fig8short_1}
	\end{figure*}
		\begin{figure*}[t!]
		\centering
		\includegraphics[width=1\linewidth]{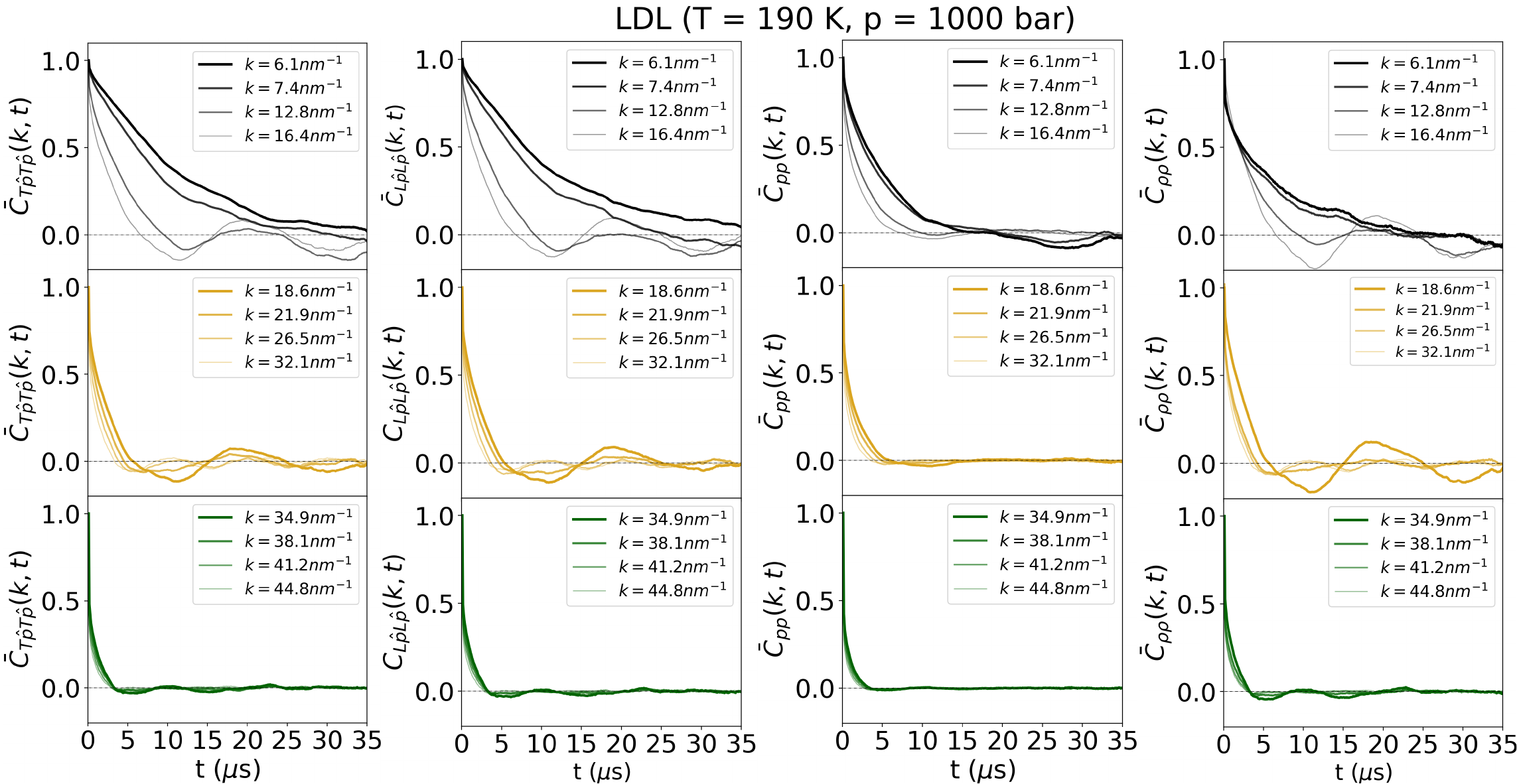}
		\caption{From left to right: $C_{T\hat{p}T\hat{p}}(k,t)$, $C_{L\hat{p}L\hat{p}}(k,t)$, $C_{PP}(k,t)$, $C_{\rho \rho}(k,t)$ at different $k$ values. All quantities are normalized to their value at $t=0$, being $\bar{C}(k,t)=C(k,t)/C(k,0)$. The oscillatory behavior in $C_{L\hat{p}L\hat{p}}(k,t)$ and $C_{T\hat{p}T\hat{p}}(k,t)$, whith characteristic frequency varying with k, indicates the presence of a collective propagating mode, consistent with the existence of a ferroelectric phase leading to the spontaneous breaking of the $O(3)$ symmetry group. Oscillations are significantly reduced in $C_{PP}(k,t)$, suggesting that the constant modulus principle is approximately satisfied. The appearance of a propagating collective mode in $C_{\rho \rho}(k,t)$ highlights an existing correlation between $P$ and $\rho$, as also emphasized in \textit{Panel b} of Fig. \ref{Fig8short_1}.}\label{Fig5short}
	\end{figure*}
	It has relevant implications on the static susceptibilities, leading to a divergence also on the static $\hat{p}$-longitudinal polarization susceptibility, $\chi_{L\hat{p}}$, for $k \rightarrow 0$ but as $k^{-1}$ \cite{Patashinskii}. Concerning dynamical correlation functions, propagating Goldstone modes are still present in $C_{T\hat{p}T\hat{p}}(k,t)$ but they are not induced in $C_{L\hat{p}L\hat{p}}(k,t)$ by the coupling between $\delta P_{T\hat{p}}$ and $P_{L\hat{p}}$ established by the constant-modulus principle. Instead, holding the constant-modulus principle, spontaneous fluctuations in $P$ generate a collective mode both in $C_{L\hat{p}L\hat{p}}(k,t)$ and $C_{T\hat{p}T\hat{p}}(k,t)$, with the same dispersion relation. Further details are provided in \textit{SI} Appendix, Sec. VI.  
	Finally, if a collective mode emerges in $C_{L\hat{p}L\hat{p}}(r,t)$, it corresponds to a mode in the density correlation function $C_{\rho \rho}(k,t)$ due to the coupling between $P$ and $\rho$, see Eq. \ref{Gibbs_ansatz}.

	The predictions above find qualitative confirmation in the analysis of MD simulations. Specifically, (i) both $\chi_{T\hat{p}}$ and $\chi_{L\hat{p}}$ in \textit{Panel a} of Fig. \ref{Fig8short_1} show a significant enhancement for $k \rightarrow 0$, approximately following $k^{-2}$ and $k^{-1}$ law, respectively. (ii) $\delta P_{L\hat{p}}$, $\delta P_{T\hat{p}}^2$ and $\delta \rho$ correlate, as emphasized in \textit{Panel b} of Fig. \ref{Fig8short_1}. (iii) Fig. \ref{Fig5short} shows that propagating modes in $C_{T\hat{p}T\hat{p}}(k,t)$, $C_{L\hat{p}L\hat{p}}(k,t)$ and $C_{\rho \rho}(k,t)$, reflecting in a their oscillatory behavior, are present. Oscillations are significantly reduced in $C_{PP}(k,t)$, showing that the constant-modulus principle is approximately satisfied.  
	The presence of a propagating mode in both $C_{T\hat{p}T\hat{p}}(k,t)$ and $C_{L\hat{p}L\hat{p}}(k,t)$, from a preliminary assessment, having in both correlation functions the same dispersion relation, linear in $k$, suggests that the observed propagating mode may originate from fluctuations in $P$. Since polarization is conserved, the
		dispersion relation of Goldstone modes is indeed expected
		to follow the $k^2$
		law, at least in the $k \rightarrow 0$ limit, as in \textit{SI}
		Appendix, Sec. VI. However, our qualitative analysis and the $k$-range accessible with current MD simulations do not allow us to exclude the possibility of coexistence between Higgs-like and Goldstone-like modes. As shown in \textit{SI} Appendix V, where different points of the ($T$, $p$) plane are analyzed, in HDL propagating modes vanish, and correlation functions decay to zero on a timescale much shorter than $\mu s$.
	\section{Conclusions}\label{sec3}
	The analysis of MD simulations from Ref. \cite{Debenedetti} alongside the construction of a classical DFT for polar liquids under mean-field approximation highlighted the role of dipolar degrees of freedom in the first-order LLPT and in the behavior around the WL.
	Consistently with a second-order ferroelectric phase transition, the theory predicts a $\chi$ divergence at the WL. A finite-size scaling analysis is needed to confirm the divergence of $\chi$ in MD simulations. The mean-field treatment proposed here is not expected, however, to describe the real critical exponents \cite{Hodge} properly . Remarkably, it was recently shown that anharmonicity in the Gibbs free energy, akin to Eq. \ref{Gibbs_ansatz}, in the ferroelectric phase can lead to $P$ fluctuations with a non-zero ensemble average. These fluctuations could reduce $P$ and, consequently, suppress the divergence of $\chi$. 
	Another hypothesis, which might explain the possible lack of $\chi$ divergence, is the occurrence of an improper ferroelectric phase transition \cite{Landau}, where the order parameter has multiple components, and the low-$T$ phase exhibits pyroelectric properties. An interesting choice for the order parameter components could be classifiers of topological order degree, introduced in Ref. \cite{Neophytou}. The lower entanglement of the hydrogen-bond network  in LDL compared to HDL \cite{Neophytou} could favor dipole alignment. Ref. \cite{Tanaka_2} suggested a two-order-parameter description for supercooled liquids.	
	
	This study disregards the potential impact of nuclear quantum effects on the LLPT in supercooled water. As discussed in Refs. \cite{Ceriotti, Wilkins} nuclear quantum effects can manifest in the reorentational dynamics of water and in the strength of hydrogen bond network, possibly being relevant for realistically locating the critical lines and CP in the $(p,T)$ plane, providing more reliable guidance for laboratory experiments. Since, furthermore, nuclear quantum effects have been observed in proton dynamics of water under electric field \cite{Cassone2}, an interesting question arising is if quantum fluctuations might reduce the ferroelectric order in the LDL and smooth the classically predicted divergence of $\chi$ at the WL, giving rise to phenomena of incipient ferroelectricity. A comparative finite-size scaling analysis between quantum and classical MD simulations of $\chi$ at the WL, in particular at high $p$'s near CP, where quantum fluctuations are expected to impact the critical behavior \cite{Vojta, Eltareb}, could shed light on this point. Nuclear quantum effects can impact also the critical exponents \cite{Vojta}. 
	
	Our analysis suggests that experimental validation of the LLPT in water can involve analyzing dielectric properties.
	It can be, furthermore, investigated whether an electric field can induce LLPT in water \cite{Wexler}, and its relationship with the $p$ and $T$-induced LLPT analyzed here.
	
	Since Pauling in 1935, attributing residual entropy in hexagonal ice to configurational proton disorder \cite{Pauling}, the concept of frustration and disorder in dipole-lattice models of ice \cite{Lasave} was introduced.
	It would be intriguing to explore whether positional and dipolar orders compete. A phase featuring dipolar order and structural disorder in LDL might correspond to one with structural order and dipolar disorder in hexagonal ice. This presumption is supported by the observation that the vanishing of the $P^2$ term in the free energy, as derived from our DFT developments, results from positional disorder.
	Finally, an alternative perspective worth considering is a glass transition in the dipolar degrees of freedom, leading to the LDL phase. This aligns with observations that $U_L$ in systems with quenched spin disorder is comparable to that with ordered spins \cite{Iniguez}. Nevertheless, possible existence of residual order must be considered, given the non-zero value of $P_i$ observed in MD simulations. If a maximum rather than a divergence were confirmed in $\chi$, it would further bolster this idea. The observed propagating collective modes in the polarization correlation functions may be linked to the breakdown of replica symmetry, possibly accompanying the dipole glass transition \cite{Parisi}.	
	
	\section{Methods}\label{sec4}
	\subsection{MD simulations of TIP4P/ice water} \label{Met1}
	The MD simulations analyzed are the same of Ref. \cite{Debenedetti}, where further details can be found. The MD simulations, performed with GROMACS, employed the classical TIP4P/Ice water model \cite{TIP4PIce} with $N=1000$ molecules in NpT ensemble, using a time-step of $2 \ fs$. This model features rigid molecule geometry with a dipole moment of $d=2.426 \ D$.  Nosé–Hoover thermostat and the isotropic Parrinello–Rahman barostat were used with characteristic time scale around respectively 10 ps and 20 ps. Electrostatic interactions were calculated by the particle-mesh Ewald method with a cutoff distance of $0.9$ nm. Van der Waals interactions has the same cutoff.
	\subsection{Analysis of MD simulations} \label{Met2}
	The vector $\textbf{P}$ at each configuration, identified by a given t, is obtained from MD simulations as follows:
	\begin{equation}
		\mathbf{P}=\sum_{i=1}^N \mathbf{d}_i, \ \ \ \        \textbf{d}_i=\sum_{\alpha} q_{\alpha}\textbf{r}_{\alpha  i}, 
	\end{equation}
	where $\alpha$ identifies the atoms of the $i$-th molecule, i.e. the two hydrogen (H), one oxygen (O) atom, and the M-site particle \cite{TIP4PIce}. The charge of the particles is $q_{\alpha}$ with $q_H=-q_M/2>0$ and $q_O=0$. $\textbf{r}_{\alpha  i}$ is the unwrapped coordinate of the $\alpha$-th particle of the $i$-th molecule. It has been verified that $d_i=d$, $\forall i$ at each configuration.
	$\chi$ and $K_T$ are respectively obtained as:
	\begin{eqnarray}
		&&\chi=\frac{<P^2>-<P>^2}{3 e_0 <V> K_B T}; \\ &&K_T=\frac{<V^2>-<V>^2}{<V>K_B T}.
	\end{eqnarray}
	$K_B$ is the Boltzmann constant and $e_0$ the vacuum permittivity. 
	The ensemble average is obtained from MD simulations by making use of the ergodicity hypothesis, thus $<O>=\bar{t}^{-1}\sum_{t=1}^{\bar{t}} O(t)$, where $O$ is a generic observable, $t=n\Delta t$ with $n \in \mathbb{N}$, $\Delta t$ is the time step and $\bar{t}$ is the time-length of MD simulations. In the present case $\Delta t \sim 8 \ 10^{-2} p s$ for all the probed points in the ($p$, $T$) plane. 
	The value of $U_L$ in a paraelectric and ferroelectric phase are established following the observations below, in analogy with the corresponding magnetic cases. 
	In the paraelectric phase, stochastic Gaussian fluctuations centered around a zero $P$ value result in $U_L \rightarrow \frac{2(n-1)}{3n}$ as $L \rightarrow \infty$ for $O(n)$ model \cite{Holm}. 
	In the present case with $n=3$, $U_L \rightarrow \frac{4}{9}$. In the ferroelectric phase, where a non-zero $P$ set in, $U_L \rightarrow \frac{2}{3}$, $\forall n$  \cite{Holm, Binder}. 
	
	The static structure factor is obtained from MD simulations as:
	\begin{eqnarray}
		&&S(\textbf{k})=<\delta \rho^*(\textbf{k})\delta \rho(\textbf{k})>,   \label{Sk}
	\end{eqnarray}   
	where the $\rho$-fluctuations $\delta \rho (\textbf{k})=\frac{1}{\sqrt{N}}\sum_{i=1}^N e^{i \textbf{k} \cdot \textbf{r}_i}$. The vector $\textbf{r}_i$ is the vector position of the center of mass of the $i$-th particle.  The symbol $^*$ states for complex conjugate. The components of \textbf{k} are derived from the expression $k_i=\frac{2 \pi}{L}$, with $i=x,y,z$. $L$ is the time-averaged simulation box size. Using instantaneous values of $L$ does not induce any substantial change. It is $L \sim 3.2$ nm, thus $\Delta k_i=2 \pi /L \sim 1.9$ nm$^{-1}$.
	To obtain $S(k)$ in \textit{SI} Appendix, Sec III, Fig. S6, averages have been taken over different $\textbf{k}$. Further averaging within a $\pm \Delta k$, with $\Delta k=1.8 \ nm^{-1}$, k-interval centered around a specific $k$ has been done to enhance the visual examination without altering the overall trends of the quantities represented. The same averaging procedure is applied to all the static and dynamic correlation functions. 
	
	The non-local static susceptibilities, transverse and longitudinal to $\textbf{k}$, are obtained from MD simulations as:
	\begin{eqnarray}
		&&\chi_{T\hat{k}(L\hat{k})}(\textbf{k})= \frac{<\delta \textbf{P}_{T\hat{k}(L\hat{k})}^*(\textbf{k})\cdot \delta \textbf{P}_{T\hat{k}(L\hat{k})}(\textbf{k})>}{ e_0 <V> K_B T},   \label{TraLongChik}
	\end{eqnarray}
	where the symbol $"\cdot "$ represents scalar product.
	It is
	\begin{eqnarray}
		&& \delta \textbf{P}_{T\hat{k}}(\textbf{k})= \hat{k} \times \delta \textbf{P}(\textbf{k}),\\
		&& \delta \textbf{P}(\textbf{k})= \sum_{i=1}^N e^{i \textbf{k} \cdot \textbf{r}_i} (\textbf{d}_i-<\textbf{d}_i>),\\
		&& \delta \textbf{P}_{L\hat{k}}(\textbf{k})=\frac{\delta \rho_c}{k} \ \hat{k},
	\end{eqnarray}
	where $\delta \rho_c=\sum_{i=1}^N \sum_{\alpha} e^{i \textbf{k} \cdot \textbf{r}_{\alpha i}} q_{\alpha}$. The definition of $\delta \textbf{P}_{L\hat{k}}(\textbf{k})$ above is preferred to $\delta \textbf{P}_{L\hat{k}}(\textbf{k})=\hat{k} \cdot \delta \textbf{P}(\textbf{k}) \ \hat{k}$, because  the latter yields less accurate results at finite $k$ due to the approximation of extended molecular dipoles as point dipoles \cite{Bopp}. 
	As for the definitions given above, $\chi_{T\hat{k}(L\hat{k})}(\textbf{k})$ are related to the space-dependent $P$ fluctuations.
	$\epsilon_{T(L)}(\textbf{k})$ are obtained from $\chi_{T\hat{k}(L\hat{k})}(\textbf{k})$ as follows: $\epsilon_T(\textbf{k})=1+\chi_{T\hat{k}}(\textbf{k})$ and $\epsilon_L(\textbf{k})=1/(1-\chi_{L\hat{k}}(\textbf{k}))$ \cite{Bopp, Dolgov}. 
	
	The non-local static susceptibilities $\chi_{T\hat{p}(L\hat{p})}(\textbf{k})$
	are obtained from MD simulations as:
	\begin{eqnarray}
		&&\chi_{T\hat{p}(L\hat{p})}(\textbf{k})=\frac{<\delta \textbf{P}_{T\hat{p}(L\hat{p})}^*(\textbf{k}) \cdot \delta \textbf{P}_{T\hat{p}(L\hat{p})}(\textbf{k})>}{e_0 <V> K_B T},   \label{TraLongChi}
	\end{eqnarray}
	It is 
	\begin{eqnarray}
		&& \delta \textbf{P}_{T\hat{p}}(\textbf{k})= \hat{p} \times \delta \bar{\textbf{P}}(\textbf{k}),\\
		&& \delta \textbf{P}_{L\hat{p}}(\textbf{k})=\hat{p} \cdot \delta \bar{\textbf{P}}(\textbf{k}) \ \hat{p} ,\\
		&& \delta \bar{\textbf{P}}(\textbf{k})= \sum_{i=1}^N e^{i \textbf{k} \cdot \textbf{r}_i} (\textbf{d}_i-\bar{\textbf{d}}),
	\end{eqnarray} 
	Strictly, it holds that $\bar{\textbf{d}}=\textbf{P}{eq}/N$ and $\hat{p}=\textbf{P}{eq}/P_{eq}$.
	In finite-size systems we approximate $\textbf{P}_{eq}$ as $<\textbf{P}>$.  It follows that the value of $\chi_{T\hat{p}(L\hat{p})}(\textbf{k}=0)$ cannot be obtained from MD simulations. Instead, we have $\delta \bar{\textbf{P}}(k=0)=\delta \textbf{P}(k=0)$ defined above. Consequently, $\chi_{T\hat{p}}(\textbf{k}=0)+\chi_{L\hat{p}}(\textbf{k}=0)=\chi$. The electric susceptibility $\chi$ remains finite in the ferroelectric phase. 
	
	The static correlations $\chi_{T\hat{p}L\hat{p}}(\textbf{k})$, $\chi_{\rho L\hat{p}}(\textbf{k})$, computed to unveil static correlation between $\delta \rho$, $\delta \textbf{P}_{L\hat{p}}$, $\delta \textbf{P}^2_{T\hat{p}}$ are obtained from MD simulations as:
	\begin{eqnarray}
		&&\chi_{T\hat{p}L\hat{p}}(\textbf{k})= \frac{|<\hat{p}\cdot\delta \textbf{P}_{L\hat{p}}^*(\textbf{k})\delta \textbf{P}_{T\hat{p}}(\textbf{k})^2>|}{[e_0 <V> K_BT]^{3/2}}; \\
		&&\chi_{\rho L\hat{p}}(\textbf{k})=\frac{|<\rho^*(\textbf{k})\hat{p} \cdot \delta \textbf{P}_{L\hat{p}}(\textbf{k})>|}{[e_0 <V> K_BT]^{1/2}}.
	\end{eqnarray}

	The dynamic correlation functions are obtained from the expressions:
	\begin{eqnarray}
		&C_{T\hat{p}T\hat{p}(L\hat{p}L\hat{p})}(\textbf{k},t)=\frac{<\delta \textbf{P}^*_{T\hat{p}(L\hat{p})} (\textbf{k},0) \cdot \delta \textbf{P}_{T\hat{p}(L\hat{p})} (\textbf{k},t)>}{e_0 <V> K_BT}; \ \ \ \   \\
		&C_{PP}(\textbf{k},t)= \frac{<\delta \textbf{P}^* (\textbf{k},0) \cdot \delta \textbf{P} (\textbf{k},t)>}{e_0 <V> K_BT}; \\
		&C_{\rho \rho}(\textbf{k},t)\propto <\delta \rho^* (\textbf{k},0) \delta \rho (\textbf{k},t)>.
	\end{eqnarray}
	They are computed with a time interval $\Delta t \sim 8 \ 10^{-2} \mu s$.
	
	\subsection{Setting the DFT of ferroelectric LLPT} \label{Met3}
	The free energy of a reference system without dipole interaction, $F_0$, and the extra free energy term, $\mathcal{F}$, accounting for dipole interaction in mean field approximation, i.e. neglecting contribution of $\tilde{\rho}(\textbf{r},\hat{d})$ correlations \cite{Hansen} are, respectively, 
	\begin{equation}
		F_0=\phi_0+T \int d \textbf{r} \ d\Omega \ \rho(\textbf{r})  \zeta(\textbf{r},\Omega) \ln[4\pi \zeta(\textbf{r},\Omega)];  \label{F0}
	\end{equation}
	\begin{equation}
		\mathcal{F}=\frac{1}{2}\int \int d \textbf{r} \ d \Omega \ d \textbf{r}' \ d \Omega' \ \rho(\textbf{r}) \zeta(\textbf{r},\Omega) w_p(\textbf{r},\textbf{r}',\Omega,\Omega') \rho(\textbf{r}') \zeta(\textbf{r}',\Omega'), \label{extraF}
	\end{equation}
	where $d \Omega$ is the infinitesimal element of solid angle. The integral is extended to the whole solid angle.
	To streamline the notation we assume here $K_B=1$. $\phi_0$ arises from the internal energy of the reference system along with the entropy term of the positional degrees of freedom. 
	The second term on the right-hand side of Eq. \ref{F0} represents the dipole orientational entropy \cite{Groh}. Though non-interacting, the dipoles are still present in the reference system.  
	The dipole interaction in Eq. \ref{extraF} is 
	\begin{equation}
		w_p(\textbf{r},\textbf{r}',\Omega,\Omega')=-\frac{d^2}{R^3}[3 (\hat{d}(\textbf{r})\cdot \hat{R})(\hat{d}(\textbf{r}')\cdot \hat{R}) -\hat{d}(\textbf{r}) \cdot \hat{d}(\textbf{r}')], \label{dipoleinter}
	\end{equation}
	where $\textbf{R}=R\hat{R}=\textbf{r}-\textbf{r}'$, $\hat{d}(\textbf{r})$ is a unit vector at the point $\textbf{r}$.
	Because spatial homogeneity assumption, $\tilde{\rho}(\textbf{r},\hat{d})=\tilde{\rho}(\hat{d})$. 
	To solve the mean-field DFT model, specifically, to find the equilibrium value of $\tilde{\rho}$ through a variational principle, we use the ansatz in Eq. \ref{ansatz_zeta} for the dipole orientation distribution,
	\begin{equation}
		\zeta(\Omega,\textbf{r})=\zeta(\Omega)=\frac{1+ \boldsymbol{\delta} \hat{d}}{4\pi}. \nonumber
	\end{equation}
	$\boldsymbol{\delta}=\delta \hat{\delta}$ satisfies $\int d\Omega \ \hat{d}(\Omega) \zeta(\Omega)= \boldsymbol{\delta}$. If $\delta \neq 0$, the polar liquid show a non-zero total polarization $\textbf{P}=d \int d\textbf{r} \rho(\textbf{r})\int \hat{d}(\Omega) d\Omega \zeta(\Omega)= d N  \boldsymbol{\delta}$, otherwise the system is isotropic and $\zeta(\Omega)$ is the uniform distribution. $\delta$ is assumed to be small, i.e. small deviations from system's isotropy are considered.
	The $\delta$-Taylor expansion around zero of $F_0$ in Eq. \ref{F0} and the computation of $\mathcal{F}$ in Eq. \ref{extraF} yields, respectively:
	\begin{eqnarray}&&F_0=\phi_0(\rho)+TN[A\delta^2+B\delta^4+B'\delta^6]; \label{F_0_ansatz} \\
		&&\mathcal{F}=- N \beta(\rho) \rho \delta^2.  \label{F_ansatz}
	\end{eqnarray}
	$A,B,B'$, which are independent of both $T$ and $p$ as $\zeta(\Omega)$ is assumed to be so, are positive constants. We expanded $F_0$ up to sixth order in $P$ since lower-order coefficients can be zero.
	To obtain Eq. \ref{F_ansatz}, due to positional disorder, a uniform distribution of the angle between $\hat{\delta}$ and $\hat{R}$ is assumed. This leads to a negative integral in Eq. \ref{extraF}, thus setting $\beta(\rho) >0$. There is no Taylor series truncation because, given the functional form of the dipolar interaction, higher order terms in $\delta$-powers are zero. Implicitly, $R \neq 0$ due to the repulsive short-range term in the reference system's interaction potential, ensuring the integral over $R$ to be finite. Note that it is the dependence of $w_p$ on $R$ to generate the density dependence of $\mathcal{F}$. Otherwise the double space integration in Eq. \ref{extraF} would simply have resulted in an $N^2$ coefficient in Eq. \ref{F_ansatz}. 
	$\rho$ slightly deviates from the reference system's density, $\bar{\rho}$, due to the slight influence of dipole interactions on it. Consequently, $\phi(\rho)$ and $\mathcal{F}(\rho)$ in Eq. \ref{F_0_ansatz} and Eq. \ref{F_ansatz}, respectively, can be given by a Taylor series in $\rho$. 
	The $\rho$-Taylor expansion around $\bar{\rho}$ of $F_0$ and $\mathcal{F}$, truncated to lowest order, leads to
	\begin{eqnarray}
		&&F_0=N\phi_0(\bar{\rho})+ M N \Delta \rho^2+TN[A\delta^2+B\delta^4+B'\delta^6]; \label{derF_0_ansatz} \\
		&&\mathcal{F}=\mathcal{F}(\bar{\rho})+\frac{\partial \mathcal{F}}{\partial \rho}|_{\bar{\rho}}\Delta \rho= - N \beta_0(\bar{\rho}) \bar{\rho} \delta^2+N\beta_1(\bar{\rho})\bar{\rho}\delta^2\Delta \rho. \ \ \ \label{derF_ansatz}
	\end{eqnarray} 
	It is $\Delta \rho=\rho-\bar{\rho}$. Since $\bar{\rho}$ is the equilibrium density of the reference system, $\phi'(\bar{\rho})=0$, and $M=\frac{1}{2N}\phi''(\bar{\rho})>0$. The single and double prime notation denotes, respectively, the first and second derivatives of $\phi$ in $\rho$. It is $\beta_0(\bar{\rho}), \beta_1(\bar{\rho})>0$. $\frac{\partial \mathcal{F}}{\partial \rho}|_{\bar{\rho}}>0$ because an increase in $\rho$, resulting in a $V$ decrease, leads to an increment of the integral in Eq. \ref{extraF}, given Eq. \ref{dipoleinter}, Eq. \ref{ansatz_zeta} and the assumption of a uniform angle distribution between $\hat{\delta}$ and $\hat{R}$ endorsed above. Finally, considering that $\boldsymbol{\delta} \propto \textbf{P}$,
	\begin{multline}
		F=\phi_0(\bar{\rho})+ M\Delta \rho^2+[TA-\beta_0(\bar{\rho}) \bar{\rho}]P^2+TBP^4+TB'P^6 + \beta_1(\bar{\rho}) P^2 \Delta \rho. \label{freeenergy_ansatz}
	\end{multline}
	The constants in Eq. \ref{freeenergy_ansatz} have been redefined, retaining, however, the same names as before.
	The thermodynamic potential in the NpT ensemble \cite{Hansen}, we want to use here, is the Gibbs free energy $G=F_0+\mathcal{F}+p N/\rho$. $\rho$ changes are induced by $V$ variations, yielding $\bar{\rho}=N/\bar{V}$ and $\Delta \rho=-\bar{\rho}\Delta V/\bar{V}$.
	Featuring an external electric field $\textbf{E}$ conjugate to $\textbf{P}$, we obtain Eq. \ref{Gibbs_ansatz}.
	The term $\gamma_0(\bar{V})$ in Eq. \ref{Gibbs_ansatz} includes the additive term $p \bar{V}$. Furthermore, it has been assumed that for a given $p$ it exists a value of $T\equiv T_c:T_cA-\beta_0 \bar{\rho}(T_c,p)=0$, see Eq. \ref{freeenergy_ansatz}. We neglect the $p$ dependence of $T_c$ for sake of simplicity. Since at $T_c$ the coefficient in front of $P^2$ becomes zero, for $T$ near $T_c$ we neglect the $T$ dependence of all the other coefficients. The $p$ dependence of $M$ is also irrelevant to our aim. Consequently, $a$, $B$, $B'$, and $M$ are positive constants. $\beta_1(\bar{\rho})$ in Eq. \ref{freeenergy_ansatz} still depends on $p$. For sake of simplicity we assume $\beta_1=\beta p$ with $\beta >0$. In Eq. \ref{Gibbs_ansatz} the factors like $1/2$ in front of $a$ have been introduced for convenience.
	\subsection{The variational principle solution for the DFT mean field model} \label{Met4}
	With the thermodynamic potential provided in Eq. \ref{Gibbs_ansatz}, the equilibrium values of $\textbf{P}$ and $\Delta V$, $\textbf{P}{eq}$ and $\Delta V{eq}$ respectively, are determined through the variational principle, establishing that at equilibrium the thermodynamic potential must be minimized, leading to Eqs. \ref{eq_PdensP6_1} and \ref{eq_PdensP6_2}, . Eq. \ref{rho} further outlines the link between $\textbf{P}{eq}$ and $\Delta V{eq}$. 
	By inserting the value of $\Delta V_{eq}$, Eq. \ref{rho}, into Eq. \ref{eq_PdensP6_1} it is obtained
	\begin{eqnarray}
		&&a(T-T_c+\frac{2\beta}{aM}p^2)\textbf{P}_{eq}+(B-\frac{2\beta^2}{M}p^2)\textbf{P}_{eq}^3+B'\textbf{P}_{eq}^5 - \textbf{E}=0. \label{Peq1}
	\end{eqnarray} 
	It is convenient to define
	\begin{eqnarray}
		&&T_c^*(p)=T_c-2\frac{\beta}{aM}p^2; \\
		&&B^*(p)=B-\frac{2\beta^2}{M}p^2. \label{b*}
	\end{eqnarray}
	$T_c^*$ is the temperature at which the term proportional to $\textbf{P}_{eq}$ in Eq. \ref{Peq1} disappears. There exists a $p$ at which $B^*=0$, causing, instead, the term proportional to $\textbf{P}_{eq}^3$ to become zero in Eq. \ref{Peq1}. We further define
	\begin{eqnarray}
		&&\bar{p}_c: B^*(\bar{p}_c)=0; \label{b*muc} \\
		&&\bar{T}_c=T_c^*(\bar{p}_c). 
	\end{eqnarray}
	The solutions of Eqs. \ref{eq_PdensP6_1} and \ref{eq_PdensP6_2} are stable if for $P=P_{eq}$ and $\Delta V=\Delta V_{eq}$, $\chi$ and $K_T$ are positive, i.e.
	\begin{eqnarray}
		\chi=\frac{\partial \textbf{P}}{\partial \textbf{E}}|_{N,T,p}>0; \ \  K_T=-\frac{1}{V}\frac{\partial V}{\partial p}|_{N,T,\textbf{E}}>0. \label{stability}
	\end{eqnarray}
	We are interested to the possible appearance of spontaneous polarization when $\textbf{E}=0$.
	From Eqs. \ref{eq_PdensP6_1}, \ref{eq_PdensP6_2} and \ref{stability}, it is found
	\begin{eqnarray}
		\chi=\frac{1}{a(T-T_c^*)+3B^*\textbf{P}_{eq}^2+5B'\textbf{P}_{eq}^4};  \label{chi_PdensP6mag0minTc}
	\end{eqnarray}
	\begin{multline}
		K_{T}=\bar{K}_{T}+\Delta K_{T} = -\frac{1}{V} \frac{\partial \bar{V}}{\partial p}+ \frac{1}{VM} \frac{a(T-T_c^*)+3B^*\textbf{P}_{eq}^2+5B'\textbf{P}_{eq}^4+4\frac{(\beta p)^2}{M}\textbf{P}_{eq}^2}{a(T-T_c^*)+3B^*\textbf{P}_{eq}^2+5B'\textbf{P}_{eq}^4}.  \label{chirho_PdensP6mag0minTc}
	\end{multline}
	$\bar{K}_{T}$ and $\Delta \bar{K}_{T}$ are implicitly defined in Eq. \ref{chirho_PdensP6mag0minTc}. It is $\bar{K}_T>0$ because $\bar{V}$ is by definition the equilibrium volume of the reference system.
	The values of $P_{eq}$, satisfying the stability conditions Eqs. \ref{stability}, are reported in Tab. \ref{tab:table4} and Tab. \ref{tab:table2} for $p$ respectively larger and smaller than $\bar{p}_c$. The corresponding value of $\Delta V_{eq}$ can be obtained from Eq. \ref{rho}. For $p>\bar{p}_c$ two stable solutions are found in the $T$ ranges specified in Tab. \ref{tab:table4}, one of which is metastable as deduced by calculating $G$ for each solution. For $p<\bar{p}_c$ only one stable solution exists for both $T>T_c^*$ and $T<T_c^*$. At a given $T<\bar{T}_c$, the difference $\Delta P^2_{eq}=P^2_{eq}(p>\bar{p}_c)-P^2{eq}(p<\bar{p}_c) \neq 0$ unless $p=\bar{p}_c$, where $B^*=0$.
	The value of $\hat{T}_c(p)$ and $\bar{T}(p)$ in Tab. \ref{tab:table4} are
	\begin{eqnarray}
		&&\hat{T}_c(p)= \frac{3}{16}\frac{{B^*(p)}^2}{aB'}+T_c^*(p);  \label{hatTc}\\
		&&\bar{T}(p)=  \frac{1}{4}\frac{{B^*(p)}^2}{aB'}+T_c^*(p) \label{Tbar}
	\end{eqnarray}
	Further details are reported in \textit{SI} Appendix, Sec. V.
	In defining a state as stable, we overlook the fact that the liquid state itself is metastable for $T$'s below the melting point.
	The behavior of $K_T$ at and along the WL is obtained in the following.
	For $T \rightarrow {T_c^*}^-$, $P_{eq}^2 \simeq \frac{a(T_c^*-T)}{B^*}$. From Eq. \ref{chirho_PdensP6mag0minTc} it is
	\begin{equation}
		\Delta K_T|_{T \rightarrow {T_c^*}^-}=\frac{1}{VM}[1+\frac{4(\beta p)^2}{MB^*}\frac{1}{2+5\frac{aB'}{B^*}(T_c^*-T)}].  \label{KTcri}
	\end{equation}
	\begin{table}[t!]
	\centering
	\caption{Equilibrium values of $P$ for the thermodynamic potential in Eq. \ref{Gibbs_ansatz} for $p > \bar{p}_c$, i.e. $B^*<0$. A dash indicates that the solution is not stable.}\label{tab:table4}
	\begin{tabular}{lcc} 
		$p>\bar{p}_c$ & $P^2_{eq}=$ & $P^2_{eq}=$\\
		&$-\frac{B^*}{2B'}+ \frac{1}{2B'}\sqrt{{B^*}^2-4a(T-T_c^*)B'}$  & 0 \\
		\midrule
		$T<T_c^*$ & stable & -\\
		$T_c^*<T<\hat{T}_c$ & stable & metastable\\
		$\hat{T}_c<T<\bar{T}$ & metastable & stable\\
		$T>\bar{T}$ & - & stable\\
		\bottomrule
	\end{tabular}
\end{table}
\begin{table}[t!]
	\centering
	\caption{Equilibrium values of $P$ for the thermodynamic potential in Eq. \ref{Gibbs_ansatz} for $p < \bar{p}_c$, i.e. $B^*>0$. A dash indicates that the solution is not stable. } \label{tab:table2}
	\begin{tabular}{lcc} 
		$p<\bar{p}_c$  & $P^2_{eq}=$ & $P^2_{eq}=$\\
		&   $-\frac{B^*}{2B'}+\frac{1}{2B'}\sqrt{{B^*}^2-4a(T-T_c^*)B'}$  & 0 \\
		\midrule
		$T>T_c^*$ \ \ \ \ \   &\ \  \ \ - &  \ \ \ \ \ \ \ \ \  stable\\
		$T<T_c^*$ & \ \ \ \ \ stable & \ \ \ \ \ \ \ \ \ - \\
		\bottomrule 
	\end{tabular}
\end{table}
	As $T \rightarrow {T_c^*}^+$, from Eq. \ref{chirho_PdensP6mag0minTc}, we find $\Delta K_T=\frac{1}{VM}$, indicating an increase in $\Delta K_T$ due to the $V$ reduction caused by the temperature decrease. As $T \rightarrow {T_c^*}^-$, according to Eq. \ref{KTcri}, $\Delta K_T$ also increases. Furthermore, Eq. \ref{KTcri} demonstrates that $\Delta K_T$ remains finite at $T_c^*$, where it reaches a maximum, as inferred from the observations above. 
	Along the curve $T=T_c^*$, where $\Delta K_T$ is maximum, $\Delta K_T|_{T =T_c^*}=\frac{1}{VM}[1+\frac{4(\beta p)^2}{MB^*}]$. Considering the definition of $B^*$ in Eq. \ref{b*} and of $\bar{p}_c$ in Eq. \ref{b*muc}, it emerges that $\Delta K_T|_{T =T_c^*}$ increases by moving along the curve $T=T_c^*$ by increasing $p$ until it diverges at $\bar{p}_c$. We can assume $\bar{K}_T$ to be constant in a sufficiently small neighborhood of $T_c^*$.
	\section{Extended Data} 
\begin{figure*} [ht]
		\includegraphics[width=1.0\textwidth]{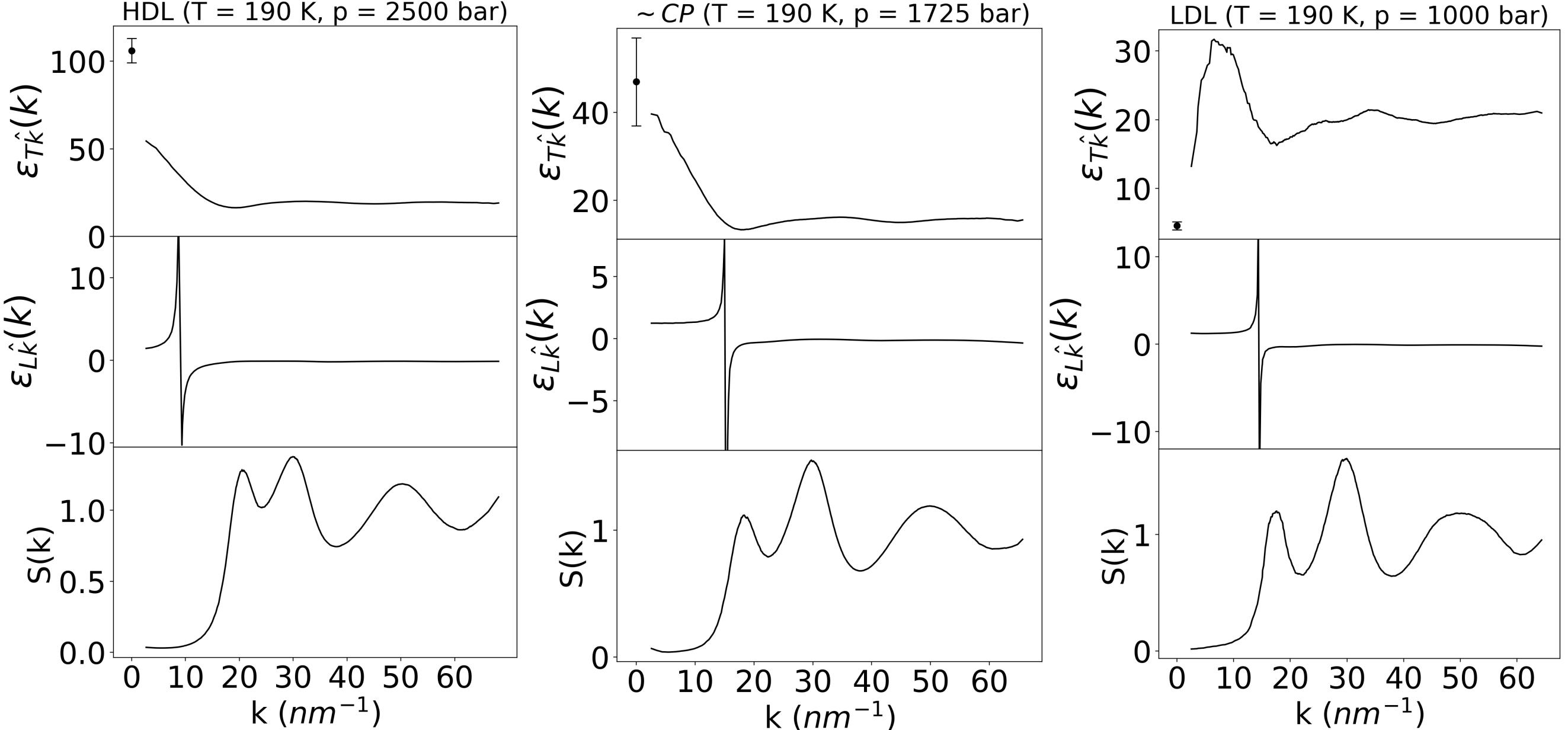}
		\caption{Non-local transverse (longitudinal) static dielectric functions, $\epsilon_{T(L)}(k)$ and static structure factor $S(k)$ in HDL, close to CP and in LDL. The black circle with the error bar, obtained by block averaging, in the upper graphs mark the value of $\epsilon_0$. The first peak in $S(k)$ corresponds to a minimum in $\epsilon_T(k)$, highlighting a possible link between the spatial arrangement of molecules' center of mass and dipole orientation.} \label{Fig7short}
	\end{figure*}

Fig. \ref{Fig7short} shows the $\textbf{k}$-dependent transverse and longitudinal (to $\textbf{k}$) static dielectric functions, respectively $\epsilon_{T\hat{k}}(k)$ and $\epsilon_{L\hat{k}}(k)$ and the static structure factor $S(k)$ in the HDL, LDL and close to $CP$. 
Since $S(k \rightarrow 0)= \rho K_B T K_T$, at CP $S(k \rightarrow 0)$ should diverge. In Fig. \ref{Fig7short} it is observed a gradual rise in $S(k)$ for small $k$ values at the thermodynamic point close to CP. A more pronounced increase, indicative of a divergence as $k \rightarrow 0$, is observed at a $k$-scale smaller than what is attainable in the current MD simulations with $N=1000$ \cite{Debenedetti}.
For $k \neq 0$, $\epsilon_T(k)$ holds only the physical meaning of a static correlation function, preserving $\epsilon_T(k)-1$ the quality of a response function only in the limit $k \rightarrow 0$, while $\chi_L(k)=1/(1-\epsilon_L(k))$ is a static response function \cite{Dolgov}. This implies that for a thermodynamically stable system, $\epsilon_L(k) \geq 1$ or $\epsilon_L(k)<0$ \cite{Dolgov, Bopp}. Additionally, a divergence occurs when $\epsilon_L=1$, as observed in Fig. \ref{Fig7short}. Another divergence has been detected at higher $k$ values under ambient conditions \cite{Bopp}, ensuring that the correct physical limit, $\lim_{k \rightarrow \infty} \epsilon_L(k) = 1$, is reached \cite{Bopp}.

	\section*{Acknowledgments} 
	The authors acknowledge Francesco Sciortino for providing the MD simulations data and engaging in insightful discussions. The authors thank Enzo Marinari, Paolo Pegolo and Achille Giacometti  for thoughtful discussions. M. G. I. acknowledges support from Ministero Istruzione Università Ricerca – Progetti di Rilevante Interesse Nazionale “Deeping our understanding of the Liquid-Liquid transition in supercooled water” COD. 2022JWAF7Y.  
	\section*{Author Contribution}
	M. G. I. conceived the idea that ferroelectricity can play a role in the liquid-liquid phase transition. M. G. I., G. P., J. R. discussed, designed and projected research. M. G. I. performed MD simulations data analysis, interpretation and theoretical developments.  M. G. I., G. P., J. R. discussed the results. M. G. I. wrote the paper. G. P., J. R. revised the paper.
	\newpage

\end{document}